\documentclass[aps,prd,twocolumn,nofootinbib,superscriptaddress]{revtex4-2}

\usepackage{amsmath,amssymb}
\usepackage{graphicx}
\usepackage{booktabs}
\usepackage{siunitx}
\usepackage{hyperref}
\usepackage{enumitem}
\usepackage[utf8]{inputenc}
\usepackage{textcomp}
\DeclareUnicodeCharacter{0327}{\c{}}
\usepackage{textcomp}

\begin{document}

\title{Probing Sub-MeV Dark Matter with Neutron-Capture $\gamma$ Spectroscopy}

\author{Bernhard Meirose}
\affiliation{Department of Physics, Chalmers University of Technology, Gothenburg, Sweden}
\affiliation{Department of Physics, Lund University, Lund, Sweden}

\author{David Milstead}
\affiliation{Department of Physics, Stockholm University, Stockholm, Sweden}


\date{\today}

\begin{abstract}
We present a general, discovery-grade framework for searching for weakly coupled new particles emitted in nuclear de-excitation following neutron capture. Rather than relying on isolated spectral features, the method exploits correlated ``satellite-line combs'': multiple weak $\gamma$-ray lines appearing at a common energy offset $\Delta$ below known capture transitions. By combining likelihood information across many parent lines and multiple target nuclei, the approach strongly suppresses nuclear-structure ambiguities and instrumental artifacts. We also discuss optimal target selection and practical experimental implementation with high-resolution HPGe detectors.
\end{abstract}

\maketitle

\section{Physics Motivation and Signature}

Light dark-sector particles in the keV–MeV mass range arise in a wide class of extensions of the Standard Model, including well-motivated dark matter scenarios such as axion-like particles, scalar mediators, and other feebly coupled states~\cite{Battaglieri:2017aum,Knapen:2017xzo,Green:2017ybv}. Compared to both weak-scale dark matter and ultralight bosonic candidates, experimental constraints in this intermediate mass window are weaker and more model dependent.

In particular, direct searches involving particles with masses in the keV-MeV range coupled to nucleons are rare. Existing experiments mostly limit couplings to electrons or photons~\cite{Alexander:2016aln,Bjorken:2009mm,Essig:2011nj}. Other searches involving nuclear de-excitations, as well as reactor-based axion searches, have been performed~\cite{Moriyama:1995bz,TEXONO:2006spf}. These searches, however, are primarily focused on electromagnetic couplings and are statistically limited in their sensitivity to hadronic interactions.

Such light-mass dark particles, if coupled to nucleons, may be produced in nuclear transitions and subsequently escape detection, carrying away a small but fixed fraction of the nuclear transition energy. Such processes can arise, for example, from the emission of axion-like particles in magnetic dipole (M1) transitions~\cite{Massarczyk:2021dje} or from more generic scalar or pseudoscalar couplings to nucleons. In this work, we explore radiative neutron-capture reactions as a laboratory probe of light dark-sector particles with nucleon couplings. To our knowledge, this environment has not been systematically exploited for dark-sector particle searches. The capture process produces discrete $\gamma$-ray cascades with precisely known transition energies, enabling correlated missing-energy searches that probe nucleon-level couplings directly.

\section{Satellite-Line Signature and Comb Method}

We examine the possibility of the emission of a light, weakly interacting particle $X$, prior to the $\gamma$-de-excitation of a compound nucleus formed in the neutron capture reaction. If the emission of the $X$-particle is a two-body process from the capture state, the energy of the $\gamma$-transition is reduced by a constant value $\Delta$. Schematically,
\begin{equation}
n + A \to B^{*}(E_x) \to B^{*}(E_x - \Delta) + X \to B(E_f) + \gamma_{\rm sat},
\end{equation}
where the satellite $\gamma$ energy satisfies
\begin{equation}
E_{\rm sat} \simeq E_0 - \Delta,
\end{equation}

where $E_0$ is the corresponding standard capture line without the presence of $X$ emission. Near the threshold, the offset $\Delta$ is approximately equal to the mass of the emitted particle, with small corrections due to nuclear recoil effects that are negligible at the energies of interest.

A fundamental aspect of this hypothesis is that the offset $\Delta$ is \emph{independent of the final nuclear state}.
In other words, for a given target nucleus, for any strong parent capture line at an energy given by $E_0^{(k)}$, there will be a weaker satellite line at an energy given by $E_0^{(k)} - \Delta$. The satellite lines will display a characteristic "satellite line comb" feature in the $\gamma$ spectrum, with constant separation between the lines and a known position relative to the parent lines.

Such a characteristic correlation will identify the signal from any conventional nuclear background. Unknown nuclear $\gamma$ lines may appear at any energy, but they will not have the characteristic correlation with a common $\Delta$ associated with many unrelated parent transitions of the same nucleus. Other possible instrumental effects, like peak tailing, coincidence summing, or escape peaks, will also be highly energy-dependent but will not have the characteristic $\Delta$ associated with the satellite lines.

In order to increase confidence that the result is not due to a particular nucleus, the same analysis will be performed on multiple target nuclei with unrelated level schemes. Although a characteristic satellite pattern may appear in a single nucleus, a statistical limit may arise due to the number of strong parent lines and background fluctuations. By performing the same measurement on multiple target nuclei, the method has discovery-level sensitivity. A true particle signal will have a characteristic energy offset $\Delta$ among different targets, while statistical fluctuations and instrumental effects will be strongly suppressed.

The underlying reason the multi-target strategy is so powerful is rooted in the physics of neutron capture rather than in statistical coincidence alone. Different target nuclei populate the compound nucleus at distinct neutron separation energies and exhibit unrelated capture cascades, leading to widely different sets of primary $\gamma$-ray energies. A real satellite signal emitted in a two-body transition near the capture state produces lines shifted by a common energy offset $\Delta$, independent of the detailed nuclear level scheme. No known nuclear process generates such a correlated offset across unrelated nuclei. This way, what is traditionally a complication in capture spectroscopy becomes a central strength of the method: target-specific features are automatically rejected, while a genuine signal must recur at the same $\Delta$ across all targets.


\section{Statistical Analysis Framework}
\label{stats}

The satellite line search is performed as a two-stage likelihood analysis,
taking into account the experimental observables and detector response.
The strategy is based on the fact that a true signal will have weak $\gamma$-ray satellites located at a common energy offset $\Delta$ with respect to a number of otherwise unrelated parent capture lines.

Before presenting the individual stages of the analysis, we summarize the experimental scales and analysis strategies relevant to the likelihood and the division of the analysis into two stages.

\subsection{Relevant Experimental Scales and Analysis Choices}

To illustrate the typical energy scales, consider a typical
neutron capture $\gamma$ line with energy $E_0 \simeq 558~\mathrm{keV}$, as
characteristic of cadmium isotopes. With a high-purity germanium detector having a given energy resolution $\mathrm{FWHM} \simeq 1~\mathrm{keV}$, corresponding to a Gaussian width
$\sigma_{\rm det} \simeq 0.42~\mathrm{keV}$, a bin width of $\sim 0.2~\mathrm{keV}$ is small compared to the detector response.

An analysis window of $\pm 20~\mathrm{keV}$ centered on each parent line--large
compared to detector resolution but small compared to nuclear transitions--has
$\mathcal{O}(10^2)$ bins and thus offers sufficient sideband
data for background characterization.

In a strong capture line with $\gtrsim 10^{4}$ detected counts, a standard
Gaussian peak fit to the data determines the centroid position $E_0$ to a
statistical accuracy of the order
\begin{equation}
\delta \mu_0 \sim \frac{\sigma_{\rm det}}{\sqrt{N}}
\sim 10\text{--}50~\mathrm{eV},
\end{equation}
while the intensity of the line is determined to a percent level or better.

Over this small analysis window, the background varies only at energy scales much
larger than both the detector resolution and the scanned offset $\Delta$. This separation of scales motivates treating the parent-line parameters as effectively fixed inputs in the subsequent satellite search.

\subsection{Stage I: Parent-Line Fits}

In the first stage, the individual parent capture lines are analyzed separately
in a region of interest.

These values, for each energy bin, are assumed to be Poisson-distributed,
including the contributions of the parametric peak shape model and the smooth
local background.

The peak model is the sum of a Gaussian function and an exponential
function, the latter modeling the charge collection tail, i.e., the
incomplete energy deposition.

This stage estimates the best-fit centroid $\mu_0^{(k)}$, the total intensity
$A_0^{(k)}$ (total counts under the peak), as well as the background
parameters, i.e., the tail fraction $f$ (fraction of peak counts under the
exponential tail) and the flat background $b$ (counts/bin).

Additionally, the analysis includes a small energy shift $\delta E$ as an
input parameter, allowing the fitted centroid $\mu_0^{(k)}$ to vary by a small
amount $\delta E$ relative to the transition energy.

No satellite contribution is assumed for the analysis of the parent lines.
The fitted values for the parent lines, i.e., $\mu_0^{(k)}, A_0^{(k)}$,
are taken as given input for the satellite analysis, as presented below.
The analysis is thus conditional on the parent transitions being already
characterized by conventional $\gamma$-spectroscopy methods, focusing exclusively on the search for satellite contributions.


\subsection{Stage II: Fixed-Offset Satellite Scan}

In the second stage, the satellite hypothesis is tested at a given energy offset
$\Delta$. For a given target nucleus, all the parent transitions are analyzed at once
under the constraint that the satellite intensity from the parent $k$ is
proportional to the measured parent intensity $A_0^{(k)}$,

\begin{equation}
A_{\rm sat}^{(k)} = f_t\,A_0^{(k)}\,R_\varepsilon(E_0^{(k)},\Delta),
\end{equation}
where $f_t \ge 0$ is a target-specific satellite fraction and
$R_\varepsilon$ accounts for the relative detection efficiency between the
parent and satellite energies.

The target-specific satellite fraction $f_t \ge 0$ is the only free parameter
in this equation, and all the parameters of the parent transitions are set to
their best-fit values from Stage~I analysis. The profile likelihood ratio test statistic is used to test the null hypothesis $f_t = 0$ at each offset $\Delta$ and is given by
\begin{equation}
q(\Delta) = -2\ln\frac{\mathcal L(f_t=0)}{\mathcal L(\hat f_t)},
\end{equation}
where the estimator $\hat f_t$ is constrained to the physical region $f_t \ge 0$. This definition follows the standard one-sided profile likelihood ratio for a parameter at the boundary of the parameter space.

The scan over $\Delta$ is carried out on a discrete grid.
Given the strong correlation for hypotheses separated by less than the detector
energy resolution, the grid is chosen to be comparable to the characteristic
resolution scale. This is performed to avoid over-sampling the highly correlated hypotheses
while adequately resolving the location of the likelihood maximum.
The background level is treated as an externally specified constant.
This background level is expected for a given energy bin, and it is
the same for all values of $\Delta$. Correlations between adjacent hypotheses in $\Delta$ arise only from the
detector energy resolution and the overlap of the parent and satellite
lines.

An actual signal is revealed as a statistically significant increase in
$q(\Delta)$ at a constant offset for multiple parent transitions and, 
importantly, for multiple target nuclei. 

To quantify the significance of such a combined excess, the total test
statistic is constructed as the sum of the individual profile likelihood
ratios for each target. In the absence of parameter boundaries, this
combined statistic would asymptotically follow a $\chi^2$ distribution
with a number of degrees of freedom equal to the number of targets.
However, due to the non-negativity constraint $f_t \ge 0$ and the scan
over $\Delta$, these asymptotic results do not directly apply. Global
significances are therefore evaluated using toy Monte Carlo simulations
of the full analysis.

\section{Robustness of the Comb Signature}
In this section, we check whether any known experimental or nuclear related effects can produce false-positive results that satisfy the fixed offset criterion imposed in the analysis.
This discussion will be limited to effects that can persist through the full likelihood scan in $\Delta$ and, hence, directly test the satellite line hypothesis.

\subsection{Instrumental Effects and Statistical Interpretation}
Instrumental effects on the shape of the individual $\gamma$-ray peaks are explicitly taken into account. The low-energy tailing is treated in the Stage I parent line fits through the inclusion of an exponential component to the core Gaussian function, where the centroid and shape parameters account for the effect. The single and double escape peaks lie at absolute energies determined by the pair production threshold, and hence they are misaligned relative to the scanned relative offset $\Delta$. Possible remaining energy scale uncertainties are addressed through small shifts of the centroid of the peaks in Stage I, which are uncorrelated between transitions. Hence, there is no effect that would give rise to an enhancement of the intensity of several parent lines at the same $\Delta$.

This profile likelihood scan in $\Delta$ constitutes a correlated set of hypotheses, where the null hypothesis is that the satellite strength is zero. Due to the non-negativity constraint on the satellite intensity, the null hypothesis is a boundary point in the parameter space. Consequently, the profile likelihood ratio does not follow the standard asymptotic $\chi^2$ distribution as dictated by Wilks' theorem~\cite{Cowan:2010js}. Global significances, which include the look-elsewhere effect, are derived using toy Monte Carlo simulations on the entire $\Delta$ scan. Discovery requires a consistent enhancement in the test statistic at the same $\Delta$ in all targets, independent of the modeling of the peak shapes, the parameterization of the backgrounds, and the modeling of the energy resolution.




\section{Target Groups for Satellite-Line Searches}

\begin{table}[h]
\centering
\caption{Representative target nuclei used in the satellite-line comb analysis. Listed are prominent parent $\gamma$-ray energies and practical notes on target preparation.}
\label{tab:targets}
\begin{tabular}{cccc}
\toprule
Group & Target & Parent energy [keV] & Comment \\
\midrule
1 & $^{197}$Au & 215  & low-energy parents \\
1 & $^{58}$Fe  & 287  & thin enriched foil \\
1 & $^{113}$In & 312  & low-energy parents \\
\midrule
2 & $^{113}$Cd & 558  & mid-energy parents \\
2 & $^{35}$Cl  & 1165 & NaCl/KCl pellet \\
2 & $^{1}$H    & 2223 & calibration line \\ 
\midrule
3 & $^{28}$Si  & 3539 & high-energy parents \\
3 & $^{56}$Fe  & 7631 & steel or foil \\
3 & $^{58}$Ni  & 8999 & foil \\
\bottomrule
\end{tabular}
\end{table}

The targets are grouped into three groups based on the energy range, referred to as Target Group 1-3, unrelated to the stages described in Sec.~\ref{stats}. The grouping covers the entire energy range of interest and allows the search to probe satellite offsets from the keV scale to several tens of keV, corresponding to particle masses accessible in neutron capture processes.

Target Group 1 is centered around low-energy parents and offers the best resolution for small offsets where the satellites appear near the parent. Sensitivity in this group is limited by the detector's resolution and low-energy tailing.

Target Group 2 covers the range of intermediate parent energies and offers a balance between large offsets and good statistics. The hydrogen capture line at 2223~keV plays a special role and is used to provide high statistics to monitor calibration and stability. Hydrogen has no strong parent lines and thus does not contribute to the search for correlated satellites. However, it provides an explicit check for any apparent satellite feature due to instrumental effects or instabilities in the energy scale: if satellites appear due to such reasons, they would also appear in the hydrogen line.

Target Group 3 contains high energy parents and is crucial in determining whether a chosen $\Delta$ value holds across vastly different energy scales. In addition, increased Compton scattering and escape peaks must be correctly modeled in this group, whose targets allow probing higher particle masses.

These target groups are designed to provide complete coverage of the $1$--$100~\mathrm{keV}$ offset scan, which directly relates to searching for new particles of mass $m_X \simeq \Delta$. Offsets lower than the keV scale are not experimentally achievable with HPGe detectors due to finite energy resolution and peak tailing at low energies, while the highest mass that can be achieved is kinematically limited by the energy of the parent capture transition.

\section{Why Correlated Satellite Searches Have Been Missed}
\label{whymissed}
Previous experiments in neutron capture $\gamma$-ray spectroscopy have been primarily concerned with the detection of individual transitions, the determination of absolute cross sections, and the measurement of angular correlations. Sensitivity to exotic effects was limited to the presence of additional, isolated peak structures, typically constrained to a level of $10^{-3}$ to $10^{-4}$ relative to strong parent transitions.

The satellite line comb structure is fundamentally different. It arises from a set of correlated excesses across many individual, otherwise statistically insignificant transitions. In a standard analysis, these are generally included in background models, attributed to unresolved structure, or dismissed as statistical fluctuations. Without a multi-line analysis based on the offset method, the defining correlation of the signal is never tested.

We can estimate an upper limit on the exotic emission fraction using a simulated neutron-capture setup with different target thicknesses. Here, $\mathrm{BR}$ is defined as the effective branching fraction for an exotic two-body emission channel competing with the standard radiative capture transition, i.e., the fraction of neutron-capture events in which a new, weakly coupled particle is emitted prior to $\gamma$ de-excitation. For simplicity, we assume a uniform neutron flux of $\phi_n = 6.81\times 10^9~\mathrm{n/cm^2/s}$ for all targets, corresponding to the measured thermal-equivalent flux at the sample position in a cold neutron PGAA beam \cite{Paul2015}. For thin foils and small samples, flux attenuation is negligible, providing a conservative estimate of signal rates. The exposure time is $T = 4$~hours, typical of historical neutron-capture $\gamma$-spectroscopy measurements, with a gamma detection efficiency of $\epsilon = 0.5\%$. The targets consist of hydrogen, gold, indium, iron, chlorine, and cadmium, all with their natural isotopic abundances and thermal neutron capture cross sections taken from evaluated nuclear data~\cite{Nobre:2025xlg}. For each isotope, the expected number of parent gamma events is computed as
\[
N_\mathrm{parent} = \phi_n \, T \, \frac{\rho}{A_\mathrm{molar}} N_A \, \sigma_\mathrm{cap} \, L \, f_\mathrm{iso} \, \epsilon ,
\]
where \(\rho\) is isotope density, \(A_\mathrm{molar}\) is molar mass, \(N_A\) is Avogadro's number, \(L\) is target thickness, \(\sigma_\mathrm{cap}\) is the thermal neutron capture cross section, and \(f_\mathrm{iso}\) is isotopic fraction. Satellite-line events were injected with a branching ratio \(\mathrm{BR}\) relative to the parent line. Per-nucleus profile likelihood scans were carried out for \(\Delta = 12 \pm 8 \text{ keV}\). For the rate estimates, a target thickness of 0.1 cm for H, \(10^{-4}\) cm for Au, 0.05 cm for In, 0.2 cm for Fe, 0.3 cm for Cl, and 0.02 cm for Cd was used. Using global significance per nucleus and LEE corrections, we obtained that \(\mathrm{BR} = 10^{-5}\) yields a maximum global significance below \(3\sigma\) for all isotopes. This does not imply that previous experiments were sensitive to such a signal, as correlated satellite-line searches were not conducted. Instead, it shows that even if the full comb-based analysis framework had been retroactively applied to the datasets with flux and exposure times typical of previous experiments, the statistics would have been insufficient to support a discovery at this branching level.

\section{Global Sensitivity and Discovery Potential}  
\begin{figure}[htb]
\centering
\includegraphics[width=\linewidth]{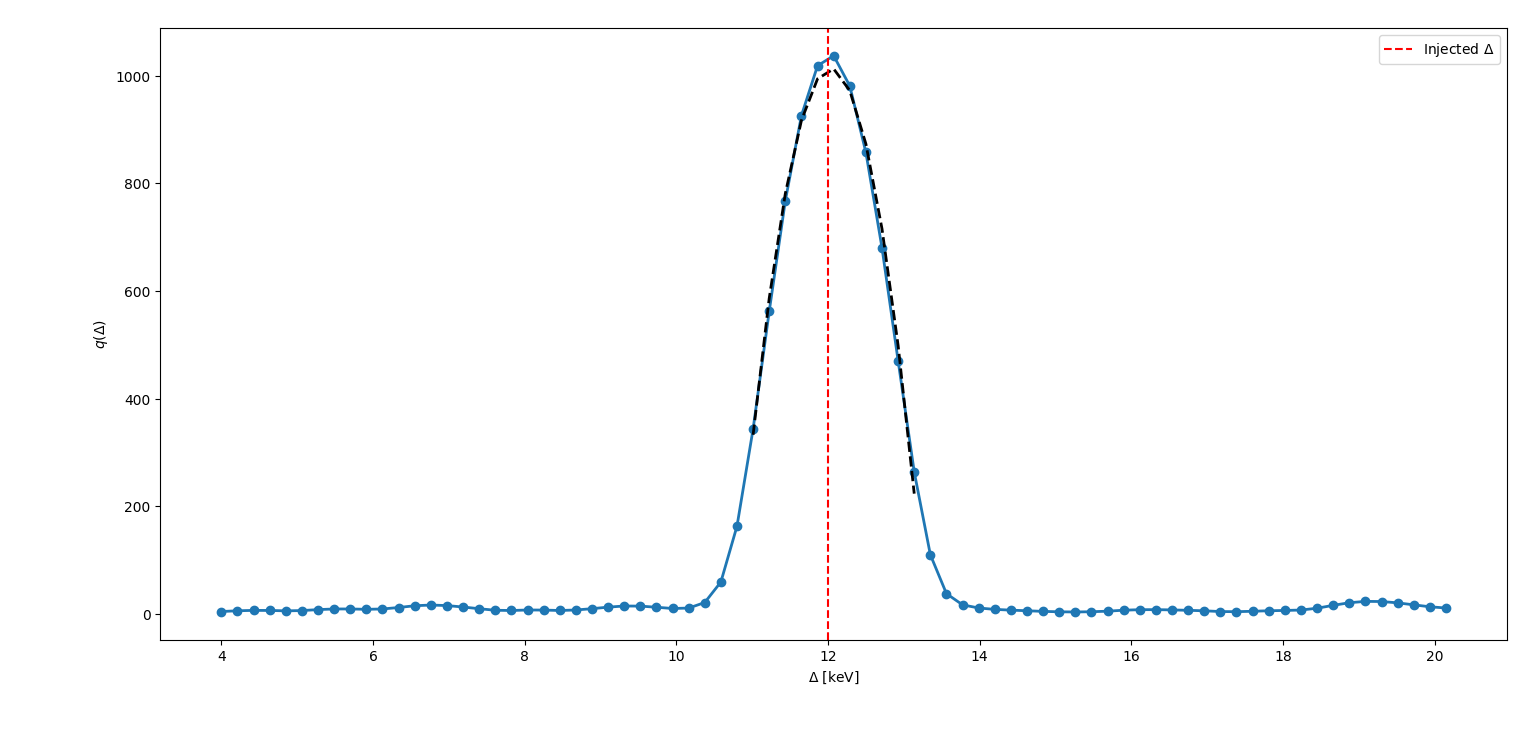} 
\caption{Profile likelihood scan $q(\Delta)$ for the combined neutron-capture targets. The maximum $q$ corresponds to the recovered $\Delta$, indicated by the dashed red line. The black dashed line shows a parabolic fit around the peak to extract $\Delta$ and the associated uncertainty.}
\label{fig:combined_qscan}
\end{figure}

Using the procedure specified in Sec.~\ref{stats}, a combined profile likelihood scan $q(\Delta)$ over all simulated neutron capture targets was performed by summing up the individual contributions from the parent lines to arrive at the total test statistic. The simulation used a Gaussian shape for the parent and satellite lines, with a width $\sigma_{\mathrm{det}} \simeq 0.42~\mathrm{keV}$ (FWHM $\sim 1~\mathrm{keV}$), compatible with HPGe detectors used for PGAA experiments. The scan was performed over 600 bins for $\pm 20~\mathrm{keV}$ around each parent line, with the resulting scan for the combined statistic depicted in Fig.~\ref{fig:combined_qscan}. Around the maximum, a parabolic fit was applied to obtain the best fit $\Delta_\mathrm{fit}$ and the associated uncertainty. The simulation setup, including the choice of target nuclei and the injected branching ratio, follows the assumptions described in Section \ref{whymissed}. The values of $q(\Delta)$ shown are local test statistics and are not corrected for the look-elsewhere effect; global significances are obtained separately using toy Monte Carlo simulations.

In a realistic analysis, the parent-line centroid, resolution, background, and normalization would be obtained from a standard local $\gamma$-spectroscopy fit and treated as fixed inputs in the satellite-line search (Stage I). In the simplified implementation used here, we do not simulate Stage I at all: the centroid, detector response, and background are held fixed at their nominal values, while only the parent normalization is re-fitted independently for each transition under the null hypothesis. This allows for the profiling of the statistical fluctuations in the parent intensity without impacting the sensitivity of the correlated satellite hypotheses.

The discovery potential depends on the total number of neutron capture events and, hence, on the product of the neutron flux and exposure time. To show the discovery potential of the method under realistic experimental conditions, we consider a typical high-flux cold-neutron beamline. To determine the LEE-corrected global significance, 300 toy data sets were simulated under the null hypothesis. For each toy data set, the combined $q_{\mathrm{max}}$ was calculated, from which the LEE-corrected global significance $Z_{\mathrm{global}}$ was derived. With a thermal-equivalent neutron flux of $\phi_{n} \sim 10^{12}~\mathrm{n/cm^2/s}$, a high-intensity cold-neutron flux achievable by the HIBEAM experiment at the ESS \cite{Santoro:2023izd}, and an exotic emission fraction of $\mathrm{BR}=10^{-8}$, the combined significance exceeds $5\sigma$, showing that correlated satellite line signatures considering such small branching fractions could be measured. 

\subsection{Impact of Detector Energy Resolution}
\label{subsec:resolution}
The sensitivity and statistical properties of the satellite line comb search are strongly dependent on the energy resolution of the detector. This is particularly important for small values of $\Delta$, as satellite peaks are then close to the relevant parent peaks.

For an ideal detector with negligible energy resolution, satellite peaks at $E^{(k)}_0-\Delta$ are well separated from parent peaks for all positive values of $\Delta$. The profile likelihood function $q(\Delta)$ will then have a well-behaved peak around the true value of $\Delta$. However, in a real experiment, the energy resolution of the detector will cause the Gaussian core and the low-energy tail of each parent peak to spill into the satellite region. When the satellite and parent peaks are separated by a value of $\Delta$ comparable to a few times the detector energy resolution $\sigma_{\text{det}}$, the satellite and parent peak shapes become partially degenerate.

\begin{figure}[htb]
\centering
\includegraphics[width=\linewidth]{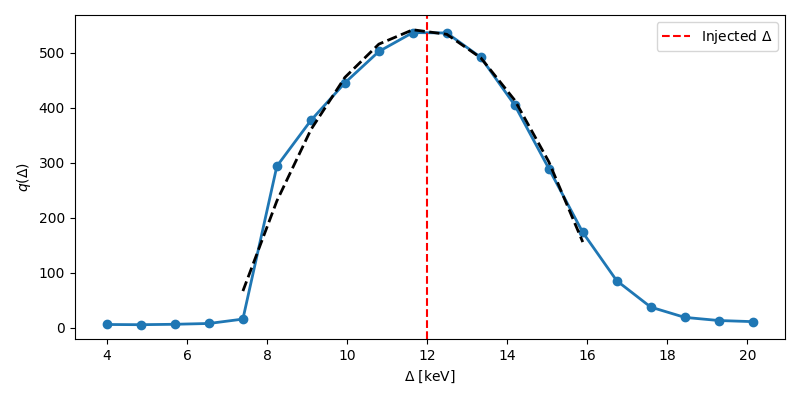}
\caption{
Combined profile-likelihood scan $q(\Delta)$ obtained assuming a detector
resolution of $\sigma_{\rm det}\simeq 4~\mathrm{keV}$.
For sufficiently small offsets $\Delta$, the satellite line overlaps the
smeared low-energy tail of the parent peak, leading to a partial degeneracy
between the satellite contribution and the parent normalization.
}
\label{fig:qscan_2kev}
\end{figure}

Figure~\ref{fig:qscan_2kev} shows the profile likelihood scan for the model~$q(\Delta)$ under the assumption of a detector energy resolution of $\mathrm{FWHM} \simeq 4~\mathrm{keV}$. The profile likelihood scan for this model displays a strong asymmetry. Figure~\ref{fig:combined_qscan} presents the profile likelihood scan for the model~$q(\Delta)$ under the assumption of a detector energy resolution of $\mathrm{FWHM} \simeq 1~\mathrm{keV}$, for which the likelihood scan is smooth and parabolic around the maximum. The strong suppression for small values of $\Delta$ results from a degeneracy between the contributions of the satellites and the contribution of the smeared low energy tail of the parent peak. For sufficiently small values of $\Delta$, the apparent excess can be compensated for by the parent normalization, while the fitted satellites will be driven toward their physical boundary at zero due to the non-negativity constraint. If this constraint were not present, the degeneracy would be absorbed by an unphysical negative contribution from the satellites line.

Note that this distortion happens only for small values of $\Delta \lesssim (3$--$4)\sigma_{\rm det}$ and does not create a false localized maximum. Instead, it creates a region of reduced sensitivity. For larger values of $\Delta$, where the satellite line is well-separated from the parent tail, the likelihood function behaves regularly, and the measured $\Delta$ remains unbiased. Therefore, this resolution effect is not a limitation of the satellite-line comb method itself. Rather, it is a result of the detector response function together with the physical constraint that the satellite intensity must be non-negative. In fact, this effect offers a clear and quantitative criterion for the minimum resolvable offset $\Delta$. Moreover, it motivates the use of high-resolution HPGe detectors for searches with small $\Delta$. Most importantly, this effect does not mimic a signal and cannot create a spurious preferred offset for multiple nuclei.

\vspace{1em}  

\subsection{Pile-up and rate considerations}
\label{pileup}

In high statistics $\gamma$-ray spectroscopy, random pile-up can occur, where two unrelated capture events take place within the pulse processing time of the detector and are recorded as a single event. The probability of this type of pile-up for a detector with shaping time $\tau$ scales approximately as $P_{\mathrm{pu}} \simeq R\tau$, where $R$ is the rate of events per crystal.

Modern digital HPGe detectors have shaping times $\tau$ of $\mathcal{O}(1-5\,\mu\mathrm{s})$, while maintaining excellent energy resolution. For per-crystal event rates of $R \lesssim 1-5\,\mathrm{kHz}$, which can be achieved with appropriate detector segmentation, this yields a pile-up probability of $P_{\mathrm{pu}} \lesssim 0.5-2.5\%$ prior to the application of pile-up rejection techniques. Digital pulse-shape analysis further suppresses unresolved pile-up contributions.

Notably, random pile-up generates a broad, rate-dependent high-energy continuum due to random energy summation and does not generate distinct secondary peaks at a fixed energy offset from the primary peaks. In contrast, the signal is comprised of correlated satellite peaks, and their energy offset from the parent peaks is solely determined by kinematics, resulting in a reproducible and target-independent fractional offset across multiple transitions and nuclei. Accidental pile-up does not have this characteristic structure and cannot replicate the multi-feature comb-like signature.

True coincidence summing within one capture cascade is also well understood and depends deterministically on detector geometry and efficiency. It can be modeled and corrected for with standard techniques. Furthermore, coincidence summing changes the peak intensities and adds sum peaks at energies corresponding to combinations of the cascade gamma-ray photons, rather than systematically shifting the peaks of individual gamma-ray lines.

The sensitivity of the proposed search is therefore not fundamentally limited by pile-up effects, but rather by the statistical limit that can be achieved under controlled count rate conditions. An optimal operating mode can be selected in which the statistical gain at higher rates is balanced against the small and correctable pile-up fraction, so that the correlated satellite signature can be resolved at the targeted branching ratios.

\section{Opportunities in existing $\gamma$-spectroscopy data.}

The method we propose can, in principle, be applied to existing
high-resolution neutron-capture $\gamma$-ray datasets, in particular,
extensive collections of spectra accumulated over many years using
HPGe detectors at high-intensity neutron sources. One notable example
is the ANNRI spectrometer \cite{ANNRI} at J-PARC, which, using
an array of HPGe detectors in a high-intensity spallation neutron beam,
has collected high-statistics capture $\gamma$-ray spectra for many
nuclei across multiple epithermal neutron resonances in the eV--keV
range. These measurements span more than a decade of operation and
represent a substantial body of data.

In addition, thermal neutron-capture $\gamma$-ray spectra are archived
in evaluated nuclear data libraries and international databases,
providing further material for retrospective studies. Complementary
datasets have also been obtained at reactor-based facilities such as
the PF1B cold-neutron beamline at the Institut Laue-Langevin, where
the EXILL array, a high-efficiency configuration of anti-Compton-shielded
HPGe detectors~\cite{Jentschel:2017sfy} was operated for nuclear
structure studies, with representative results reported in
Ref.~\cite{Scheck:2025rqz}. These measurements cover a broad energy range
(from tens of keV up to several MeV) and include coincidence-resolved
$\gamma$-ray cascades from neutron capture and fission reactions.
While primarily optimized for $\gamma$--$\gamma$ correlation studies
of complex level schemes, such datasets may also be reanalyzed in terms
of single-line spectra for correlated satellite-line searches.

In the absence of a clear signal in existing datasets, a particularly
promising opportunity for a dedicated search is provided by the planned
HIBEAM experiment \cite{Santoro:2023izd} at the European Spallation Source.

\section{Generality of the Search}
It is important to note that the satellite signature is not unique to
neutron capture. Other processes, such as radioactive decay or fusion evaporation,
could, in principle, also produce a signal if an additional weak channel were present.
These, however, offer either lower statistics or less structured spectra, which limits their discovery potential for weak satellite signals. Neutron-capture, therefore, is the process that maximizes sensitivity to very small branching fractions.

The choice of detector technology presents a complementary trade-off.
HPGe detector arrays with high efficiency offer the statistical power
required for discovery-mode searches by maximizing event rates for
many transitions. In contrast, crystal diffraction $\gamma$-ray spectrometers,
such as the GAMS\cite{HGBorner_1993}, can achieve orders-of-magnitude better energy
resolution (eV scale) for some transitions, providing greater sensitivity to smaller energy offsets $\Delta$. However, their low efficiency and limited angular coverage limit their reach for extremely weak branching ratios. A natural experimental strategy is to use high-efficiency detectors as discovery tools and then to use ultra-high-resolution detectors to characterize weak signals identified by the high-efficiency detectors.

Finally, we note that in high-flux neutron environments, HPGe detectors are subject to
progressive radiation damage, which degrades energy resolution and
induces non-Gaussian line-shape asymmetries over time. These effects,
observed in long-running capture experiments, must be monitored and
incorporated into the empirical line-shape modeling of Stage~I. They do not,
however, introduce correlated features at a fixed offset $\Delta$, and
therefore cannot mimic the satellite-line comb signature.

\section{Conclusion}
We have introduced a discovery-oriented framework for searching for weakly coupled dark sector particles in neutron capture gamma-ray spectra. The approach takes advantage of the unique signature of correlations between satellite lines and different parent transitions and targets, converting what is typically a source of complexity in nuclear structure into a tool for background rejection and discrimination.
The approach of using multiple targets aims to achieve sensitivity over a vast range of particle masses, from a few keV up to several MeV, and is shown through basic calculations to be able to access extremely weak branching ratios, well beyond what is currently feasible.
The satellite line comb represents a new approach in the exploration of sub-MeV physics in the context of dark sector particles and is complementary to existing approaches. It provides a clear path forward for future high-resolution and high-sensitivity experiments.

\section*{Acknowledgments}
The authors thank Mike Snow for helpful discussions and for pointing out the ANNRI neutron-capture $\gamma$-ray datasets at the Japan Proton Accelerator Research Complex. The authors thank Andreas Heinz for helpful discussions.

\bibliographystyle{apsrev4-2}
\bibliography{refs}  

\end{document}